\documentclass{article}
\usepackage{arxiv}
\usepackage[utf8]{inputenc} 
\usepackage[T1]{fontenc}    
\usepackage{hyperref}       
\usepackage{url}            
\usepackage{booktabs}       
\usepackage{amsfonts}       
\usepackage{nicefrac}       
\usepackage{microtype}      
\usepackage{lipsum}		
\usepackage{graphicx}
\usepackage{mathtools}
\usepackage{doi}

\title{Computation of Reliability Statistics for Finite Samples of Success-Failure Experiments}

\date{May 6, 2023}	

\author{ \href{https://orcid.org/0000-0002-9700-0749}{\includegraphics[scale=0.06]{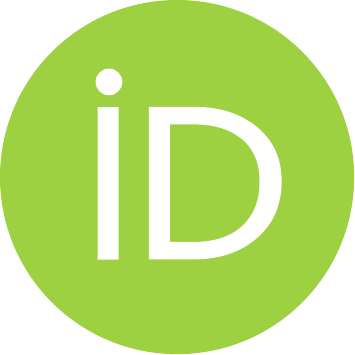}\hspace{1mm}Sanjay M.~Joshi}\thanks{https://www.linkedin.com/in/sanjaymjoshi/} \\
	Independent Researcher \\
	\texttt{sanjaymjoshi@iitbombay.org} \\
}

\renewcommand{\shorttitle}{Reliability Statistics for Finite Samples of Bernoulli Trials}

\hypersetup{
pdftitle={Reliability Statistics for Finite Samples of Bernoulli Trials},
pdfsubject={stat.ME, math.ST, stat.AP, stat.CO},
pdfauthor={Sanjay M.~Joshi},
pdfkeywords={Confidence, Reliability, Assurance, Sample Sizes, Bernoulli Trials},
}

\begin{document}
\maketitle

\begin{abstract}
Computational method for statistical measures of reliability, confidence, and assurance are available for infinite population size. If the population size is finite and small compared to the number of samples tested, these computational methods need to be improved for a better representation of reality. This article discusses how to compute reliability, confidence, and assurance statistics for finite number of samples. Graphs and tables are provided as examples and can be used for low number of test sample sizes. Two open-source \texttt{python} libraries are provided for computing reliability, confidence, and assurance with both infinite and finite number of samples.
\end{abstract}


\section{Introduction}
Reliability of a part, functionality, or experiment is the probability that it is successful or does not fail. Such success or failure experiments are often called Bernoulli or binomial trials. The outcome or sample is a random variable with binomial distribution. We will assume that the samples are independent of each other and are identically distributed. Computational methods of related statistical measures, Confidence and Assurance, are reviewed in \cite{joshi23b} for infinite population size.

This article presents computational methods for reliability, confidence, 
and assurance for finite number of samples. Let's quickly review the computations for infinite number of samples from \cite{joshi23b}, since these will be reused in this article.

\subsection{Confidence}

Confidence, $c$, is the probability that the actual reliability is at least $r$. If there are $f$ failures in $n$ samples, the computation is \cite{iso16269}:
\begin{equation}
    c = 1 - \sum_{k=0}^f \binom{n}{k}  (1-r)^k r^{n-k},
    \label{C_f}
\end{equation}

\subsection{Reliability}

If we want to calculate the minimum reliability $r$ for a confidence level $c$, given $f$ failures in $n$ samples, we solve equation \ref{C_f}for $r$. There is no closed-form expression for the general case, but numerical methods are available for the computations and have been implemented in \cite{joshi23}.

\subsection{Assurance}

Set assurance, $a = r = c$, in equation \ref{C_f} to get
\begin{equation}
    a = 1 - \sum_{k=0}^f \binom{n}{k}  (1-a)^k a^{n-k}
    \label{A_f}
\end{equation}

Like reliability computations, this equation can be solved using numerical methods. A solution has been implemented in \cite{joshi23}.

\section{Computations with Finite Population Size}

\subsection{Confidence}

Suppose tests of initial $n$ samples yield $f$ failures and we want to compute statistics for $m$ additional samples. The number of additional failures we can encounter is between 0 and $m$. These result in discrete steps of reliability. The statistics at the extremes are discussed first.
\begin{itemize}
    \item Suppose none of the additional samples fail. We have already encountered $f$ failures. Therefore, the reliability cannot be greater than $1 - f/(n+m)$. In other words, confidence is zero for $r > 1 - f/(n+m)$ if $f \ne 0$. If $f = 0$, confidence is zero for $r = 1$.
    \item Suppose all the additional $m$ samples fail. Even in this worst case, the reliability cannot be less than $1- (f+m)/(n+m)$. In other words, confidence is 1 for $r \le 1 - (f+m)/(n+m)$.
\end{itemize}

To compute confidence in reliability between these two extremes, let's look at the discrete steps of reliability, at $d$ failures out of the $m$ additional samples, $0 < d < m$ (We will consider the special case of $d=0$ later). The corresponding level of reliability needed to meet this requirement of failures is $r_n = 1 - d/m$. Given $f$ failures in $n$ samples, the confidence in this level of reliability can be readily computed using equation \ref{C_f}, for infinite population.

To compute confidence in reliability level other than the discrete values above, we first compute the highest discrete reliability value $1-d/m$ less than or equal to desired reliability level and use the confidence in that level of reliability.

Now let's look at the special case of $d = 0$. We cannot use the $r_n = 1 - d/m$ approach, since that will yield $r_n = 1$, which does not yield meaningful value of confidence. Therefore, we consider the case of exactly 1 failure in $m+1$ additional samples. That yields $r_n = 1 - 1/(m+1)$ and we can compute the confidence using \ref{C_f}.

To summarize, here is how the confidence in reliability for $m$ additional samples is computed given $f$ failures in $n$ samples:
\begin{itemize}
    \item $0 \le r \le 1 - (f+m)/(n+m)$: Confidence is 1.
    \item $1 - (f+m)/(n+m) < r < 1 - f/(n+m)$: Use confidence of $r_n = 1 - d/m$ using equation \ref{C_f}, where $d$, $0 < d < m$, is lowest possible value that yields $r_n \le r$.
    \item At $r = 1 - f/(n+m)$: If $ f > 0$, use confidence of $r_n = 1 - 1/(m+1)$ using equation \ref{C_f}. If $f=0$, confidence is 0.
    \item $1 - f/(n+m) < r \le 1$: Confidence is 0.
\end{itemize}

The zero failure case is probably the most commonly encountered situation. Let's see how the confidence with finite number of samples compares to that with infinite samples. With $f = 0$,
\begin{itemize}
    \item $0 \le r \le 1 - m/(n+m)$: Confidence is 1, higher than $1 - r^n$
    \item $1 - m/(n+m) < r < 1$: Use confidence of $r_n = 1 - d/m$ using equation \ref{C_f}, where $d$, $0 < d < m$, is lowest possible value that yields $r_n \le r$. Therefore, the confidence in $r_n$ is higher than or equal to the confidence in $r$.
\end{itemize}

Hence, for finite samples with zero failures, the confidence in any level of reliability is greater than or equal to the confidence in that level of reliability with infinite samples.

\subsection{Reliability}

To compute minimum reliability given confidence $c$, we simply compute the confidence $c_n$ for $r_n = 1 - d/m$, varying $d$ from $0$ to $m$. As soon as $c_n$ meets or exceeds $c$ for a certain value of $d$, we use the associated reliability $r = 1 - (f+d)/(n+m)$ as the value of reliability at the desired level of confidence $c$. 

\subsection{Assurance}

To compute assurance, we follow similar approach to reliability computations. For each value of $d$, $0 \le d \le m$, we compute reliability as $r = 1 - (f+d)/(n+m)$ and confidence $c$ using the method above. The assurance for this $d$ is $a_d = \min(r, c)$. The assurance is maximum $a_d$ for $0 \le d \le m$.

These computation methods were implemented in \texttt{python} in an open-source library \cite{joshi23}. Ready-to-use interfaces using \texttt{Jupyter} notebooks were implemented in another open-source library \cite{joshi23a}. Some sample results from these are discussed in the next section.

\section{Computational Results}

\subsection{Confidence}

\begin{figure}[htp]
	\centering
 \begin{tabular}{cc}
     \includegraphics[scale=0.5]{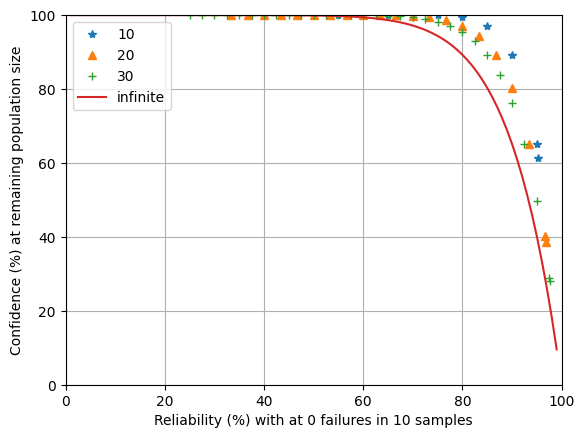} & \includegraphics[scale=0.5]{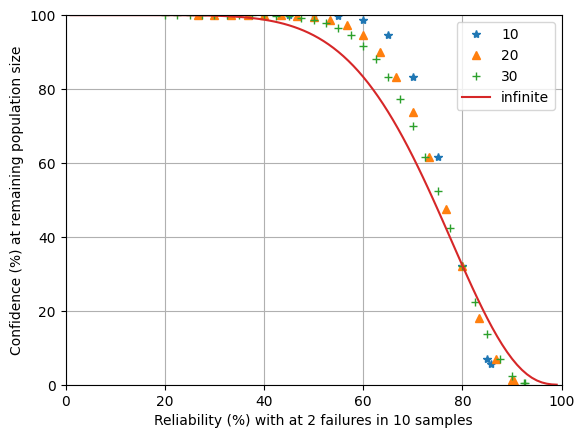} \\
      (a) $n=10$, $f=0$ & (b) $n=10$, $f=2$
 \end{tabular}
	\caption{Confidence in reliability as number of additional samples change.}
	\label{fig:conf}
\end{figure}

Figure \ref{fig:conf} shows how confidence changes with size of additional samples as desired reliability changes. Subplots 'a' and 'b' show the behavior with zero and two failures, respectively, in 10 test samples. The number of additional samples are varied from 10 to 30, and the resulting confidence at discrete levels of reliability is shown with discrete points. For reference, confidence for infinite sample size is shown with a continuous line.

With zero failures, the confidence level at a given reliability level drops as the number of additional samples increases. The ultimate lower limit is shown as the line for infinite samples. This is expected behavior, as discussed above. The number of discrete steps in reliability and the highest values change based on the number of additional samples. For example, with $m = 10$ additional samples, the two highest reliability values (shown with blue stars) are $r = 1 - 1/10 = 90\%$ corresponding to one additional failure and $r = 1 - 1/11 = 91\%$ corresponding to zero additional failures. For reliability higher than that, the plot does not show any additional points, since for all values above that level of reliability, the confidence is zero.

The confidence levels for $m = 20$ and $m = 30$ are between $m = 10$ and infinite samples, as expected. The plots for $m = 20$ and $m = 30$ end at $r = 1 - 1/21$ and $r = 1 - 1 / 31$, respectively, since confidence for any higher level of reliability is zero. The infinite sample size plot continues all the way up to the ultimate limit of 100\% reliability.

With two failures in 10 samples, interestingly, reliability level of $1 - 2/10 = 80\%$ marks the "cross-over point". For reliability less than 80\%, the confidence at a certain reliability is higher for lower number of additional samples, exactly as in the case of zero failures. At reliability more than 80\%, though, the confidence at a certain reliability is \emph{lower} for lower number of additional samples. 

As the number of additional samples decrease, the 'room' to tolerate additional failures decreases, resulting in lower confidence. With infinite additional samples, there is always some room for additional failures, hence the confidence is higher than finite number of additional samples.

For zero failure case, the cross-over point is at $1 - 0/n = 100\%$ reliability. Therefore, for all finite number of additional samples, the confidence at any discrete level of reliability is higher than that for infinite number of additional samples.

\subsection{Reliability}

\begin{figure}[htp]
	\centering
  \begin{tabular}{cc}
     \includegraphics[scale=0.5]{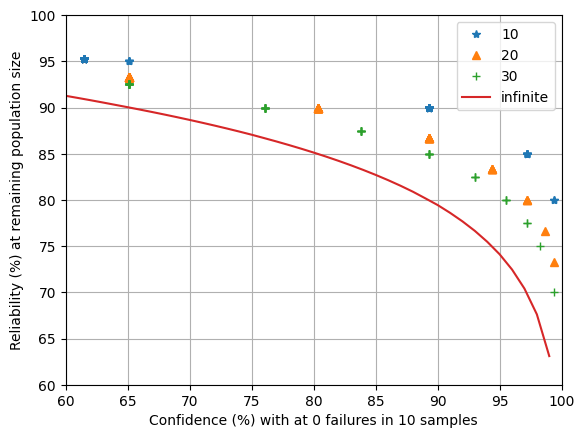} & \includegraphics[scale=0.5]{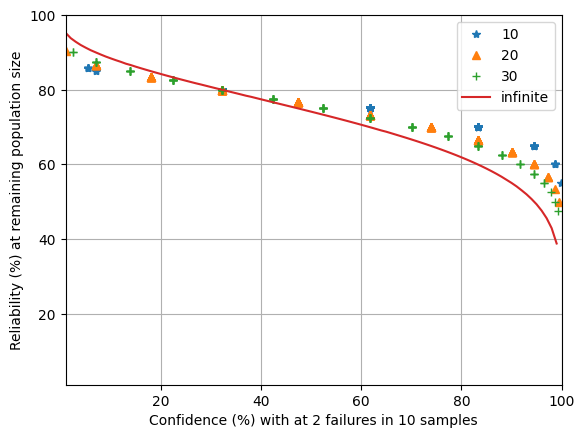} \\
      (a) $n=10$, $f=0$ & (b) $n=10$, $f=2$
 \end{tabular}
	\caption{Reliability at different levels of confidence as number of additional samples change.}
	\label{fig:reli}
\end{figure}

Figure \ref{fig:reli} how reliability changes with size of additional samples as desired confidence changes. Subplots 'a' and 'b' show the behavior with zero and two failures, respectively, in 10 test samples. As expected, these plots show similar behaviour as seen in the confidence plots seen in Fig. \ref{fig:conf}.

For easier visibility, subplot 'a' is zoomed-in to show how confidence increases with different number of additional samples with zero failures in $n=10$ samples. At about 89\% confidence level, the reliability is 90\% with 10 additional samples, 85\% with 30 additional samples, and only 80\% with infinite sample sizes.

Subplot 'b' shows the cross-over point at reliability of $1 - 2/10 = 80\%$ (on the \emph{y}-axis!) achieved at confidence of a little higher than 30\%, corresponding to two failures in 10 samples as discussed above.

The reliability plots are essentially the transposed versions of the confidence plots from the previous subsection. The shape seems different only because the axes are transposed.

\subsection{Assurance}

Table \ref{tab:assurance} shows the assurance levels at different number of additional samples (along columns) as the number of test samples increase down the row. The first entry in each row is the number of tested samples with zero failures. The number of additional samples increase from 1 to 10 in each column, with infinite samples as the last column.
 
\begin{table}[htp]
    \centering
\begin{tabular}{r|rrrrrrrrrrr}
\hline
\hline
    & \multicolumn{9}{c}{Number of additional samples} \\
    &    1 &    2 &    3 &    4 &    5 &    6 &    7 &    8 &    9 &   10 &   inf \\
\hline
  3 & 80.0 & 80.0 & 70.4 & 71.4 & 75.0 & 70.4 & 70.0 & 72.7 & 70.4 & 69.2 &  68.2 \\
  4 & 83.3 & 83.3 & 80.2 & 75.0 & 77.8 & 80.0 & 74.0 & 75.0 & 76.9 & 76.0 &  72.4 \\
  5 & 85.7 & 86.8 & 86.8 & 77.8 & 80.0 & 81.8 & 81.4 & 76.9 & 78.6 & 80.0 &  75.5 \\
  6 & 87.5 & 88.9 & 88.9 & 82.2 & 81.8 & 83.3 & 84.6 & 82.2 & 80.0 & 81.2 &  77.8 \\
  7 & 88.9 & 90.0 & 90.0 & 86.7 & 83.3 & 84.6 & 85.7 & 86.7 & 82.8 & 82.4 &  79.7 \\
  8 & 90.0 & 90.9 & 90.9 & 90.0 & 84.6 & 85.7 & 86.7 & 87.5 & 86.6 & 83.3 &  81.2 \\
  9 & 90.9 & 91.7 & 92.3 & 92.3 & 86.6 & 86.7 & 87.5 & 88.2 & 88.9 & 86.6 &  82.4 \\
 10 & 91.7 & 92.3 & 92.9 & 92.9 & 89.3 & 87.5 & 88.2 & 88.9 & 89.5 & 89.3 &  83.5 \\
 11 & 92.3 & 92.9 & 93.3 & 93.3 & 91.4 & 88.2 & 88.9 & 89.5 & 90.0 & 90.5 &  84.5 \\
 12 & 92.9 & 93.3 & 93.8 & 93.8 & 93.1 & 88.9 & 89.5 & 90.0 & 90.5 & 90.9 &  85.2 \\
 13 & 93.3 & 93.8 & 94.1 & 94.4 & 94.4 & 90.7 & 90.0 & 90.5 & 90.9 & 91.3 &  86.0 \\
 14 & 93.8 & 94.1 & 94.4 & 94.7 & 94.7 & 92.2 & 90.5 & 90.9 & 91.3 & 91.7 &  86.6 \\
 15 & 94.1 & 94.4 & 94.7 & 95.0 & 95.0 & 93.5 & 90.9 & 91.3 & 91.7 & 92.0 &  87.2 \\
 16 & 94.4 & 94.7 & 95.0 & 95.2 & 95.2 & 94.6 & 91.5 & 91.7 & 92.0 & 92.3 &  87.7 \\
 17 & 94.7 & 95.0 & 95.2 & 95.5 & 95.5 & 95.5 & 92.7 & 92.0 & 92.3 & 92.6 &  88.2 \\
 18 & 95.0 & 95.2 & 95.5 & 95.7 & 95.8 & 95.8 & 93.8 & 92.3 & 92.6 & 92.9 &  88.6 \\
 19 & 95.2 & 95.5 & 95.7 & 95.8 & 96.0 & 96.0 & 94.7 & 92.6 & 92.9 & 93.1 &  89.0 \\
 20 & 95.5 & 95.7 & 95.8 & 96.0 & 96.2 & 96.2 & 95.4 & 93.1 & 93.1 & 93.3 &  89.4 \\
 21 & 95.7 & 95.8 & 96.0 & 96.2 & 96.3 & 96.3 & 96.1 & 93.9 & 93.3 & 93.5 &  89.7 \\
 22 & 95.8 & 96.0 & 96.2 & 96.3 & 96.4 & 96.6 & 96.6 & 94.7 & 93.5 & 93.8 &  90.1 \\
\hline
\end{tabular}
\vspace{10pt}

\caption{Assurance levels for a number of additional samples (along column) at zero failures in a number of test samples (along rows)}
\label{tab:assurance}
\end{table}

Here is an example of how to use the table:
\begin{itemize}
    \item Suppose we have successfully tested 3 samples with 0 failures. If we plan to make only 5 more samples, the assurance is 75\% (as opposed to 68.2\% if we had planned to make infinite samples, shown in the last column). That is, reliability is at least 75\% with confidence of at least 75\%.
    \item Suppose this first batch of 5 more samples also yields zero failures. Now we have tested 8 samples with zero samples. If we want to make 8 more samples in the next batch, the assurance is 87.5\%.
\end{itemize}

The table shows some strange behavior, though: the assurance does not always reduce as
the number of additional samples increases! For example, in the first row for $n = 3$, The assurance increases from 71.4\% for 4 additional samples to 75\% for 5 additional samples! Shouldn't assurance decrease as we increase the number of additional samples?

Remember that for finite number of samples, assurance is the minimum of reliability and confidence. To better understand assurance, we need to look at both reliability and confidence as the number of additional samples and number of additional failures in those samples change.

\begin{figure}[htp]
	\centering

   \begin{tabular}{cc}
     \includegraphics[scale=0.5]{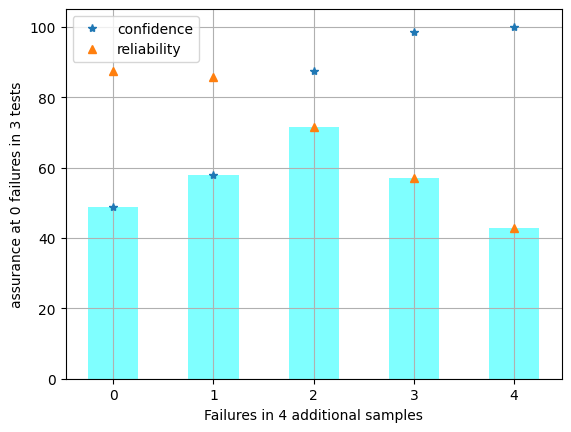} & \includegraphics[scale=0.5]{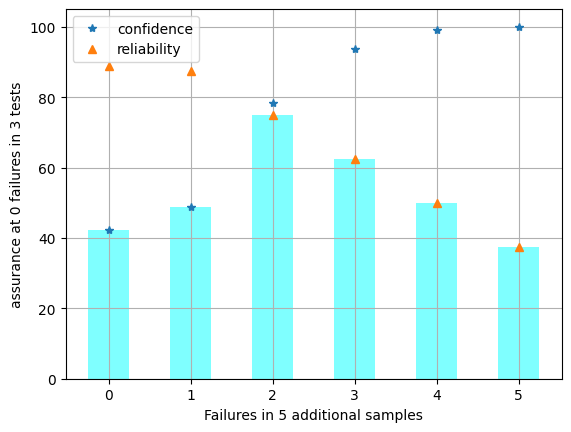} \\
      (a) $n=10$, $f=0$, $m=4$  & (b) $n=10$, $f=0$, $m=5$
 \end{tabular}
 
	\caption{Reliability, confidence, and assurance as number of additional samples change.}
	\label{fig:assur}
\end{figure}

Figure \ref{fig:assur} shows the reliability and confidence as number of failures $d$ from $m$ additional samples increase from 0 to $m$, with subplot 'a' for $m = 4$ and subplot 'b' for $m = 5$. The number of initial test samples is $n = 3$ with $f = 0$ failures.

From subplot 'a' for $m=4$, the reliability at $f+d = 2$ failures out of $3+4 = 7$ samples is $1 - (0+2)/(3+4) = 71.4\%$. For this reliability and 4 additional samples, the confidence with 0 failures in first 3 samples is 87.5\%. The assurance is minimum of these two, which is the reliability level of 71.4\%.

From subplot 'b' for $m=5$, the reliability at $f+d= 2$ failures out of $3+5 = 8$ samples, is $1 - (0+2)/(3+5) = 75\%$, a little higher. The confidence with 0 failures in first 3 samples drops to 78.4\%. The assurance, minimum of these, is still the reliability level of 75\%. This value, though, is higher than the 71.4\% that we got for $m = 4$! The much higher level of confidence at $m = 4$ did not help gain higher level of assurance.

We can think of assurance as the trade-off between the reliability and confidence. The smaller the trade-off, i.e., the closer reliability and confidence are to each other, the higher the assurance. Due to discrete values of reliability levels, we can sometimes get slightly higher assurance even if the number of additional samples increase.

\begin{figure}[htp]
	\centering
    \begin{tabular}{cc}
     \includegraphics[scale=0.5]{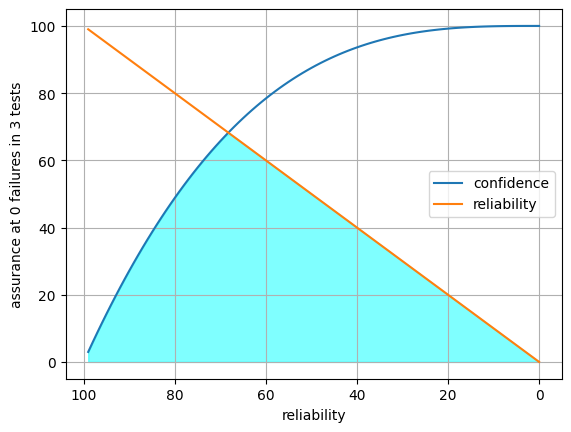} & \includegraphics[scale=0.5]{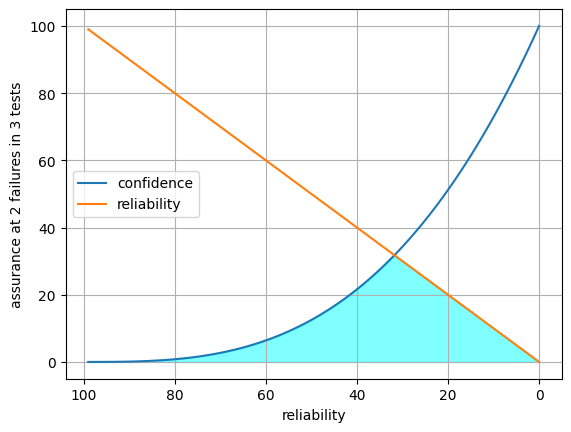} \\
      (a) $f=0$  & (b) $f=2$
 \end{tabular}
  
	\caption{Reliability, confidence, and assurance as the minimum of these two.}
	\label{fig:assur_inf}
\end{figure}

With infinite additional samples, it is possible to compute confidence at any level of reliability. Therefore, assurance occurs where the trade-off is zero, i.e., confidence is equal to reliability. We can see this behavior in Fig. \ref{fig:assur_inf}. Subplot 'a' for $n=3$ and $f=0$ shows assurance of 68.2\% at the cross-section of reliability and confidence curves. In subplot 'b', with $f=2$, the assurance at cross-section drops to only 31.8\%. Note that the x axis is inverted to keep consistency with the axes in Fig. \ref{fig:assur}.

\section{Conclusions}

This article presented computational methods for reliability, confidence, and assurance for success-failure experiment with finite population size. The basis of these computations is that the level of reliability needed in additional number of samples is less than or equal to the overall level of reliability. The confidence is then computed for the level of needed reliability in additional samples assuming infinite population size, for which computational methods have been established already. The resulting confidence is greater than or equal to that for infinite samples.

These statistical measures show a larger difference from those measures for infinite population size as the number of additional samples in the population decreases. Thus, these methods will present more meaningful information to decision makers evaluating proposals to make a small batch of initial production units based on success-failure rate of test batch.

The assurance levels do not show uniform decrease as the number of additional samples in population increases. The article discussed the impact of trade-off between reliability and confidence giving rise to such seemingly anomalous behavior. We hope the assurance table presented in this article helps decision makers plan optimal sample sizes to take advantage of higher values of assurances.

The graphs and tables are created from a \texttt{Jupyter} notebook using a new open-source \texttt{python} library. Both of these can be easily modified to obtain statistics for reliability, confidence, and assurance for finite number of samples.

\bibliographystyle{unsrtnat}


\begin{thebibliography}{1}

\bibitem{joshi23b}
      S.M. Joshi,
      \newblock  \href{https://doi.org/10.48550/arXiv.2303.03167} {"Computation of Reliability Statistics for Success-Failure Experiments"}, 
     \newblock arXiv:2303.03167 [stat.ME],
    \newblock 2023

\bibitem{iso16269}
International Organization for Standardization (ISO)
    \newblock \emph{ISO 16269-6:2014
Statistical interpretation of data — Part 6: Determination of statistical tolerance intervals},
    \newblock Ed. 2, 2014 

    \bibitem{joshi23}
    S.M. Joshi.
    \newblock \emph{Reliability Statistics Library},
    \newblock \url{https://github.com/sanjaymjoshi/relistats},
    \newblock accessed May 14, 2023.


    \bibitem{joshi23a}
    S.M. Joshi.
    \newblock \emph{Reliability Statistics Notebook},
    \newblock \url{https://github.com/sanjaymjoshi/relistats_notebook},
    \newblock accessible online as \href{https://colab.research.google.com/github/sanjaymjoshi/relistats_notebook/blob/main/relistats_binomial.ipynb} {relistats\_binomial.ipynb on Google Colab} 
    \newblock accessed May 14, 2023.

 \end{thebibliography}

\end{document}